# Post Processing Graphical User Interface for Heat Flow Visualization

Lars Olt[1], Luis Diego Fonseca Flores[2], and Ian Mckinley[3]
*Jet Propulsion Laboratory, California Institute of Technology, Pasadena, California 91109, USA*

**Thermal Desktop (TD) is an industry-standard thermal analysis tool used to create and analyze thermal models for landers, rovers, spacecraft, and instrument payloads. Currently, limited software exists to extract and visualize metrics relevant to heat flow within TD, impeding thermal engineers from analyzing their results quickly. This paper discusses a graphical user interface (GUI) built in MATLAB and C++ which uses TD's application programming interface (API), OpenTD, and a custom parser to address this void. Specifically, we present a method for efficiently loading temperature, conductance, and submodel metrics using a side effect of TD's Compressed Solution Results (CSR) files. This approach can reduce the runtime for correlating model nodes and conductors with submodel IDs by orders of magnitude. Lastly, we reflect on the shortcomings of this method for reading data, consider the future of the GUI, and provide recommendations for subsequent OpenTD releases.**

## Nomenclature

| | | |
|---|---|---|
| *TD* | = | Thermal Desktop |
| *HFV* | = | Heat Flow Visualizer |
| *NASA* | = | National Aeronautics and Space Administration |
| *JPL* | = | Jet Propulsion Laboratory |
| *API* | = | Application Programming Interface |
| *GUI* | = | Graphical User Interface |
| *CSR* | = | Compressed Solution Results |
| *PDF* | = | Portable Document Format |
| *NODTRE* | = | Node Tree Binary |
| *EMIT* | = | Surface Mineral Dust Source Investigation |
| *ELC1* | = | ExPRESS Logistics Carrier 1 |

## I. Introduction

THERMAL analysis is a fundamental component of the aerospace industry as it underpins all aerospace systems' reliability, safety, and performance. From a mission's inception through its termination, accurate thermal analysis is critical for maintaining margins of safety and meeting requirements. Components may experience overheating, material degradation, or stress without thorough thermal analysis, leading to performance deterioration and catastrophic failures.

TD is the industry standard software for simulating thermal cases. It allows engineers to model heat transfer between components of complex scenarios and determine thermal loads missions must account for. However, while modeling heat flow is a standardized process, analyzing the results of these simulations is not. Because of the diversity in missions and scenarios modeled, even within one organization, thermal teams rely on several unique scripts and programs to achieve their desired analyses. Many of these tools, including that discussed in this paper, rely on the OpenTD API for programmatic access to TD. Hume L. Peabody, based at Goddard, published a paper[1] that explores a variety of OpenTD's use cases relevant to NASA in particular, providing valuable insights into its capabilities. Peabody's paper further examines the limitations of the API and offers workarounds to them.

---

[1] Computer Science Undergraduate Student, Western Washington University, Bellingham, WA 98225.
[2] Thermal Engineer, Cryogenic Systems Engineering Group, 4800 Oak Grove Drive, Pasadena, CA 91109.
[3] Thermal Engineer, Cryogenic Systems Engineering Group, 4800 Oak Grove Drive, Pasadena, CA 91109.

The tool described in this paper, Heat Flow Visualizer (HFV), represents a continuation of Peabody's work and that of JPL engineers in efficiently post-processing large thermal scenarios. In particular, HFV is an application for thermal engineers to quickly and effectively analyze and produce summaries of their thermal cases. It was developed to provide a level of standardization for analyzing thermal scenarios, allowing engineers to customize generated visualizations to fit their particular needs.

## II. Optimizing Data Reading

TD has various export formats, allowing users to prioritize specific metrics of thermal cases for analysis. HFV uses the CSR format, as it stores a broad scope of information regarding the thermal model and natively provides transient data. Accessing data from CSR files comes with challenges, however, in terms of efficiency at scale. Two primary factors contribute to this inefficiency: the method architecture and underlying data structures. In terms of architecture, OpenTD exposes critical variables through a series of objects and methods. For instance, accessing temperature data at the submodel level involves creating a CSR dataset object and separate ItemIdentifierCollection objects for each submodel before retrieving the corresponding temperature data. The second inefficiency arises from the underlying data structures employed by OpenTD – particularly those related to the CSR dataset object as they lack thread safety. Consequently, performing multithreading operations with the API is impractical for the large-scale models that could benefit from them, limiting OpenTD's performance to single-threaded optimizations when handling data relevant to heat flow.

```cpp
// reads number of submodels in SIZES binary, from CSR directory, to numSubs pointer
int64_t* numSubs;
string sizesPath = rootDir + "SIZES";
ifstream sizeFile(sizesPath, ios::in | ios::binary);
sizeFile.read(reinterpret_cast<char *>(numSubs), sizeof(int64_t));
sizeFile.close();
```

**Figure 1. Reading from CSR Binaries.** *This code segment demonstrates reading a single value from a CSR binary into a pointer variable.*

To mitigate these challenges, a parser was developed in C++ to replace OpenTD in loading conductance and temperature data. It leverages spatial and temporal locality when reading CSR files, significantly improving runtime for submodel-related calculations. Figure 1 provides a simplified example of reading data from a CSR binary file as intuition for the process implemented in the parser. By loading all relevant data at once and capitalizing on the sequential storage of TD node indices in the NODTRE binary file, the parser achieves a notable enhancement in runtime, as seen in Figure 2. This graph represents a comparison between runtime and the number of node temperature data that are loaded for multiple TD models. For example, a model that contains ten nodes and ten transient instances would have an x-value of one hundred. In particular, the inflection in OpenTD's trendline is caused by compounding redundant data that is read into memory as part of the documented pipeline for accessing submodel-specific metrics.

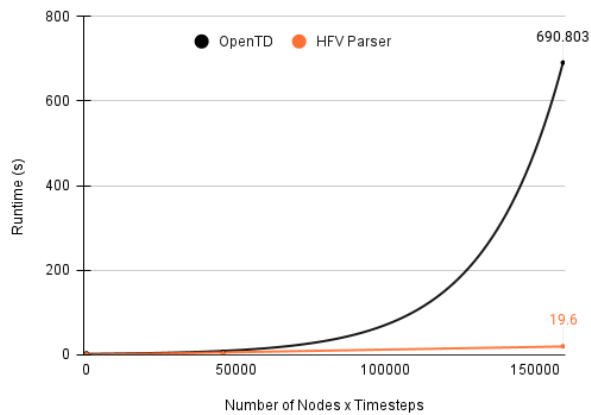

**Figure 2. Runtime Analysis, Loading Temperatures.** *File size, given as the product of the number of nodes and number of timesteps, is compared with runtime to demonstrate the advantage of parsing values with a custom implementation. Each model's runtime is averaged over five runs.*

Analyzing the complexity class difference between the two methods of loading data provides further insight into the growth rate of each respective function. OpenTD's implementation for loading temperature data is a $\Theta(n^2)$ operation, whereas the HFV parser runs in $\Theta(n)$ time – significantly reducing the runtime for large, transient thermal models.

The most notable speed enhancement achieved is a byproduct of an architectural decision in the NODTRE file within the CSR output. This file stores data in blocks comprising a head and body. The head of each block contains the submodel name and the number of node indices in the body, among other metadata. The body stores the node indices associated with the submodel. When designating model nodes to their respective submodel, the conventional approach documented by OpenTD reads all data within the NODTRE file related to the passed submodel.

In contrast, our method reads only the head of each data block, skipping the body. The values can be interpreted later because the blocks are stored sequentially based on the node indices. This storage method effectively encodes the index values into the node count stored in the head. Figure 3 provides a visual representation of the block structure in the NODTRE file, illustrating the efficiency gained through this approach.

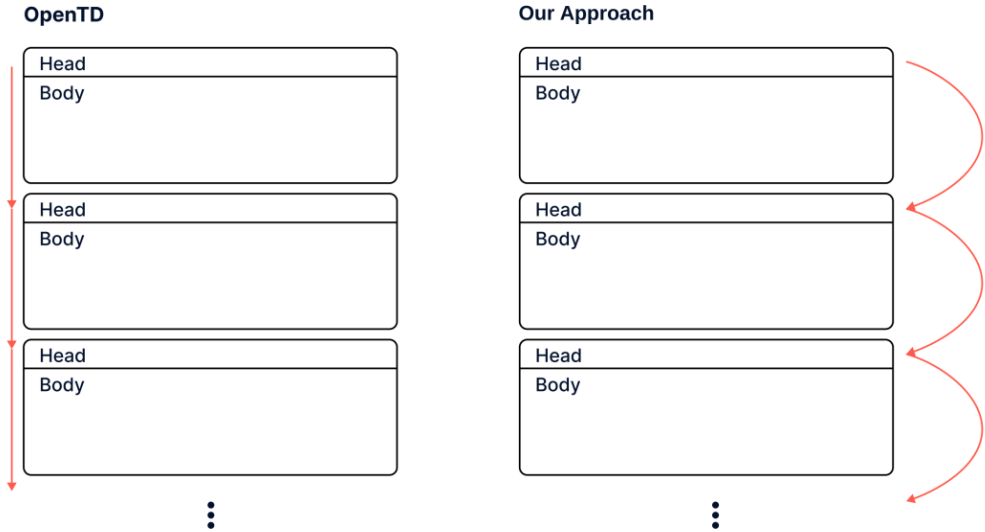

**Figure 3. Data Reading Scheme.** *Scheme implemented by the HFV parser to read values faster than OpenTD.*

It is important to note that the parser only retrieves temperature data and calculates the overall conductances between submodels. Due to time constraints during development, heat flows are tentatively calculated between submodels using OpenTD.

### III. Visualization Standard

A visualization standard was developed to present the loaded data. It was created for optimal readability and to convey information quickly. Figure 4 illustrates a sample visualization generated by the GUI.

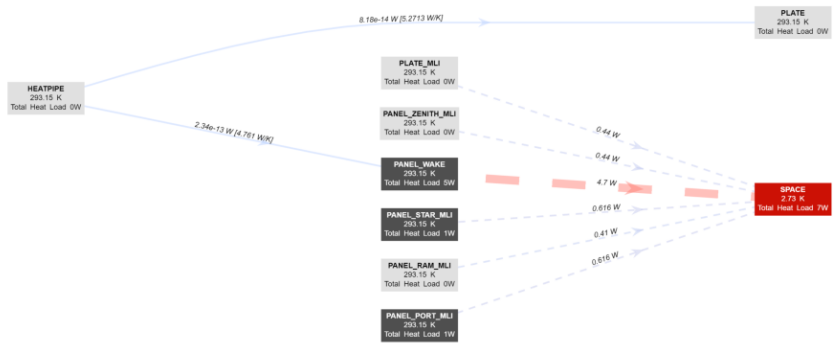

**Figure 4. Example Visualization.** *Heat flow diagram generated by HFV.*

Rectangles represent TD submodels and contain critical information such as the submodel name, average node temperature, and total heat load. The annotation for total heat load represents the difference between incoming and outgoing heat transfer. Rectangle color indicates the heat load on the respective submodel. Those with light gray backgrounds represent loads less than one absolute value, dark gray backgrounds indicate excess outgoing, and red backgrounds represent excess incoming. This distinction is made to highlight environmental loads.

Arrowed lines depict the direction of linear or radiative heat transfer between individual submodels. Solid lines symbolize heat transfer by conduction, while dashed lines signify heat transfer through radiation. The numerical value along each line denotes the heat transfer degree measured in the chosen units. Additionally, linear conduction arrow lines display their respective conductance values within encased brackets. The magnitude of heat transfer within the visualization determines all edges' color and line weight. Red lines indicate higher values than blue lines, facilitating the identification of hotspots for heat transfer and pinpointing potential heat leaks between submodels.

MATLAB's graph data type was used to create the visualizations. Doing so allows for the convenient association of weights and relationships between submodels. Furthermore, it provides native layout algorithms, made accessible to the user through a dropdown menu offering four layout options:

a. Layered layout (Figure 5a): organizes submodels into strata, aiding in the identification of heat transfer between specific submodels of interest.
b. Forced digraph layout (Figure 5b): creates a force-directed graph by assigning weights based on submodel-to-submodel heat transfer, grouping related submodels at large scale.
c. Subspace embedding layout (Figure 5c): arranges submodels from a high-degree subspace into two dimensions, facilitating the clustering of related submodels.
d. Circular layout (Figure 5d) plots the submodels in a circular arrangement, enabling heat flow analysis within defined simulation regions.

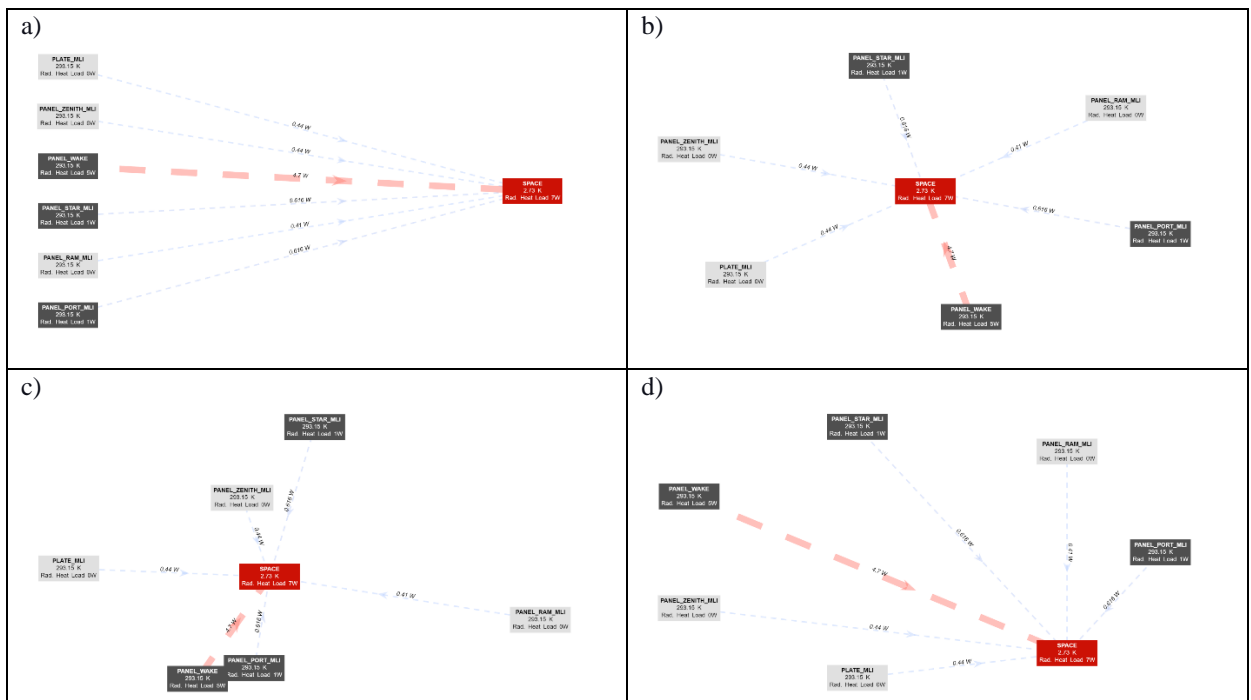

**Figure 5. Visualization Algorithms.** a: *layered, b: forced digraph, c: subspace embedding, d: circular.*

## IV. The GUI

HFV was implemented using MATLAB's App Designer. MATLAB was selected as the primary language for the GUI due to its widespread use at JPL – to promote engineer participation. As seen in the Figure 6 flow chart, once data is loaded and visualized, HFV allows the user to save the visualization in a PDF file that stores all exports for the respective project.

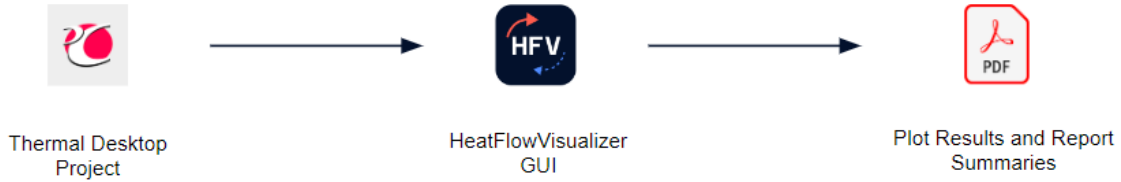

**Figure 6. GUI Usage.** *Flow chart of GUI's intended usage with Thermal Desktop.*

As an example of one use case of the GUI, Figure 7 represents the thermal model and exported heat flow visualization from a spaceflight instrument called EMIT (Earth Surface Mineral Dust Source Investigation). This instrument is mounted on the ExPRESS Logistics Carrier 1 (ELC1) platform of the International Space Station (ISS). Paired with the intuition for the underlying model, following heat flow in the generated diagram from the plate to the radiators and out to space is intuitive.

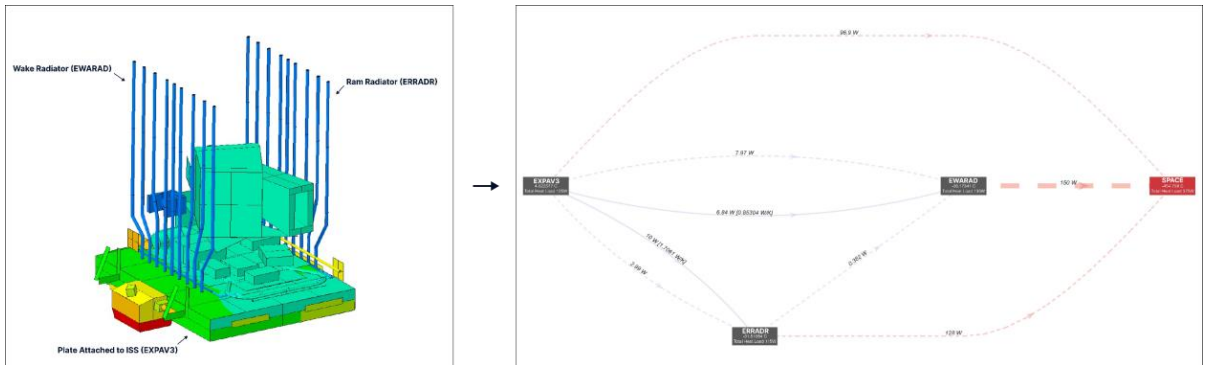

**Figure 7. EMIT Thermal Model.** *Illustration of Thermal Desktop model in AutoCAD and corresponding heat flow visualization created in HFV.*

The primary functionality of the GUI itself is to simplify thermal cases down to digestible visualizations. The toolbar at the top of the window includes features to streamline this process. The most prominent three allow the user to specify submodels to include, group submodels together, and select a low-bounds cut-off for radiant heat flow. To demonstrate, Figure 8 is an untouched thermal case visualized in HFV. Figures 9, 10, and 11 represent examples of tuning each of the abovementioned parameters individually.

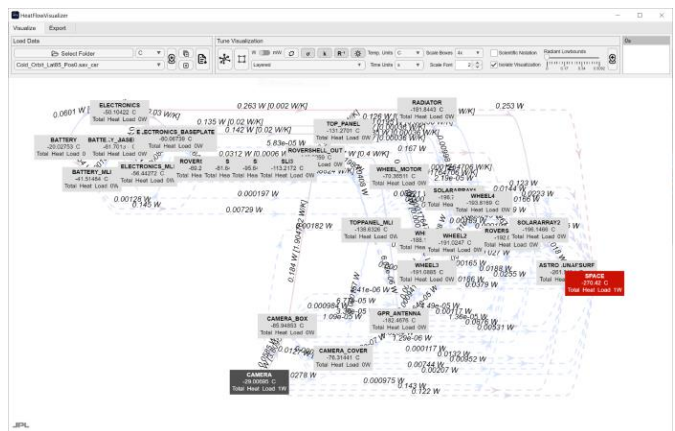

**Figure 8. Unprocessed Data in HFV.** *Example of a raw thermal case loaded into the GUI.*

Beginning with submodel selection, the user can specify which submodels to include in the visualization from a generated list of submodels. This feature helps analyze particular sections of a thermal model. Figure 9 shows that specifying submodels is the most explicit means of simplifying thermal scenarios.

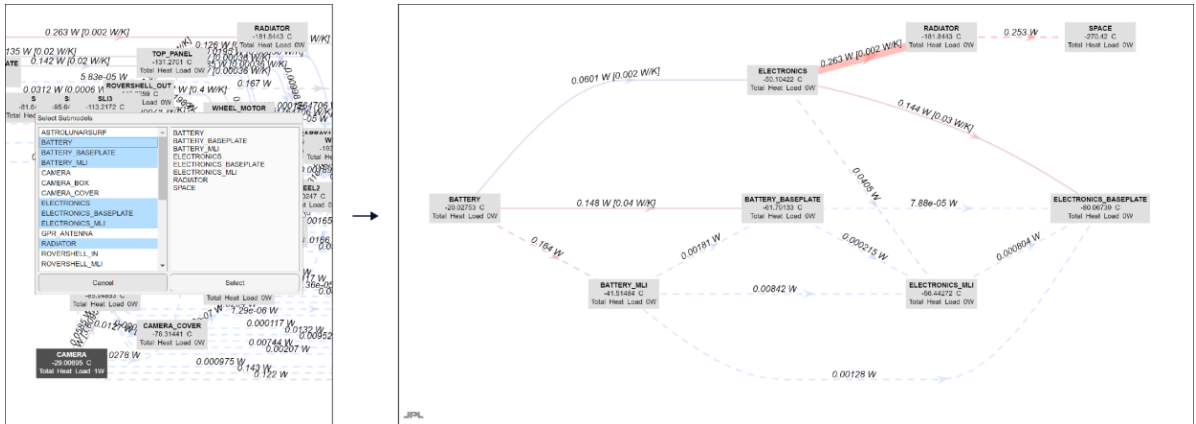

**Figure 9.  Data Processed with Submodel Selection.** *Example of the same thermal case as pictured in Figure 8, processed with submodel selection.*

In many cases, however, you will want all submodels represented in the visualization. In this situation, grouping submodels may be a more effective means of simplification. As the name suggests, the user simply adds submodels to groups, which are represented as single rectangles in the visualization. Each added rectangle receives the averaged data from all of the corresponding submodels in the group. Figure 10 shows this feature is particularly useful in cases where many submodels comprise one section of the thermal model.

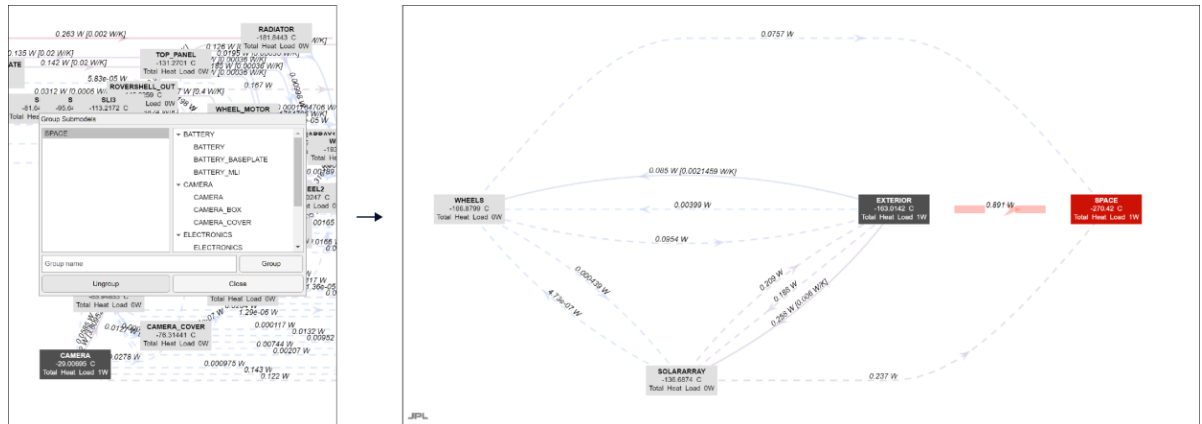

**Figure 10.   Data Processed with Grouping.** *Example of the same thermal case as pictured in Figure 8, processed with grouping.*

One of the most powerful features in HFV for simplifying thermal scenarios is the radiant heat flow lower bounds slider. As the name suggests, the user can specify a low-bounds cut-off threshold for radiant heat flow. Thus, any values below the specified slider value get omitted from the visualization. As seen in Figure 11, and as you might imagine, this feature is useful in nearly all scenarios as insignificant radiant heat slips between submodels are a common occurrence.

It is important to mention that these tools are intended to be used in unison. Figure 11 shows that overlaps in the visualization can persist after tuning just one parameter.

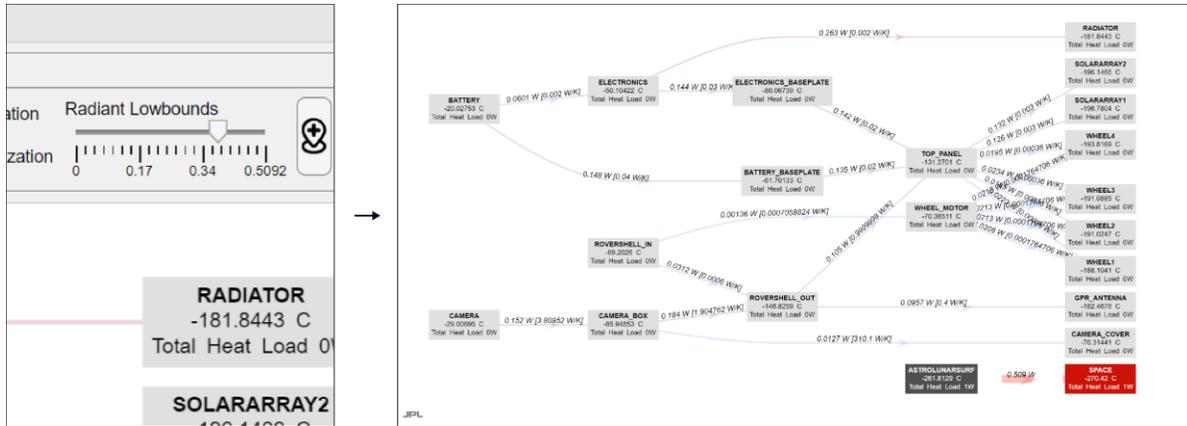
**Figure 11.   Data Processed with Heat Flow Lower bounds Slider.** *Example of the same thermal case as pictured in Figure 8, processed with the heat flow lower bounds slider.*

Finally, while the GUI primarily emphasizes generating heat flow visualizations, additional output graph are available. Specifically, the GUI allows user to graph transient submodel temperatures and submodel-to-submodel heat flow, as illustrated in Figures 12 and 13.

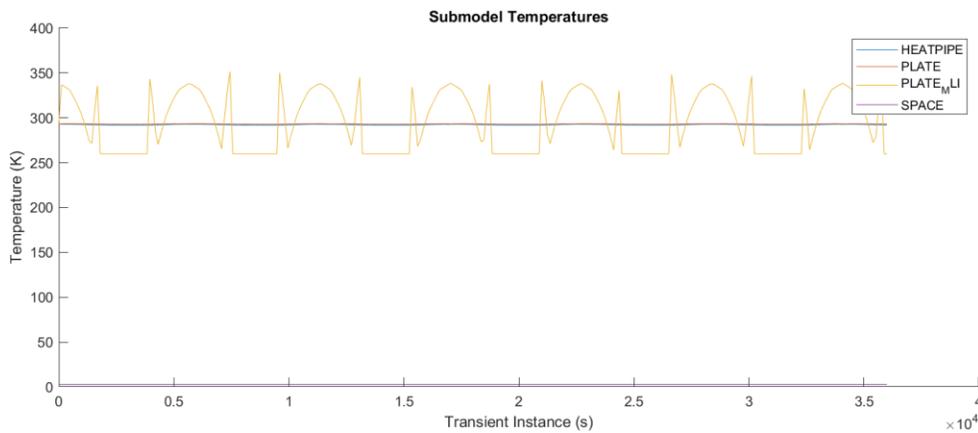
**Figure 12.   Transient Submodel Temperature Graph.** *Example of transient submodel temperature graph from HFV.*

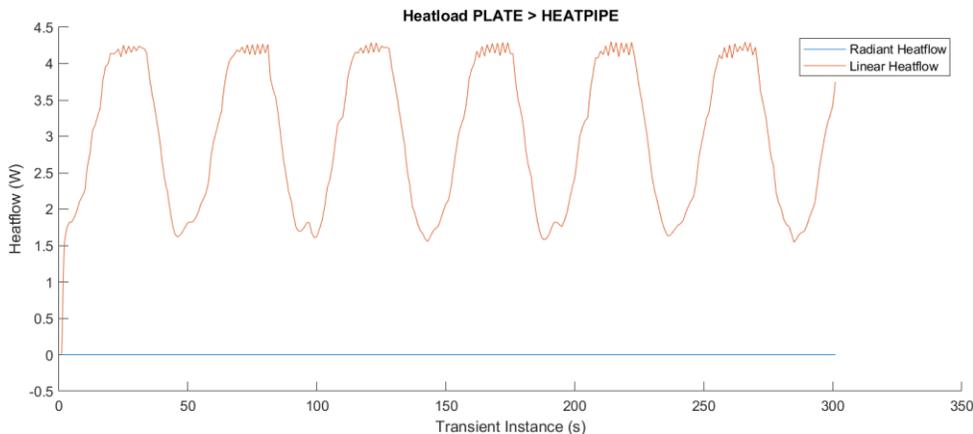
**Figure 13.   Transient Heat Flow Between Submodels.** *Example of transient heat flow between PLATE and HEATPIPE submodels from HFV.*

## V. Optimizing Data Storage

As the size of thermal files increases, so do the resources required to read the values into memory. For small TD models, the impact of this process on runtime is negligible. However, at the scale of missions such as Europa Clipper, with hundreds of submodels over hundreds of transient instances, minimizing data read is crucial for a tolerable user experience.

To reduce load time for thermal cases that were previously loaded, after the parser reads the values, HFV automatically creates a project directory and stores them in custom binary files. These files are dynamically added to as the user subsequently includes more data in their visualizations and switches between transient instances. Doing so ensures that returning to previously loaded data does not require reloading the full thermal case.

## VI. Conclusion

HFV is a valuable interface for engineers to analyze unique thermal scenarios rapidly for their projects. In this paper, an association algorithm was presented that significantly accelerates data loading, surpassing the speed of OpenTD. Additionally, a heat flow visualization format was introduced along with tuning parameters, offering customization to meet users' specific needs. Finally, the diverse graphical output options available through HFV were demonstrated.

It is important to emphasize that HFV relies on OpenTD to calculate heat transfer between submodels. Additionally, while the binary files created by the GUI minimize resource usage for loading data, deleting these files has not yet been implemented and must be done manually. Regarding the visualizations themselves, submodels are currently not yet user-draggable. These issues will be addressed in future versions of the tool, reflecting our commitment to refining its functionality and ensuring accurate and comprehensive heat flow analysis.

Regarding accessing HFV, it is tentatively restricted to use within NASA. However, distributing the GUI will be considered should external interest present itself.

Efficient programmatic access to data is crucial for applications that regularly process large and complex thermal scenarios. While the parser outlined in this paper addresses this need now, it is not a definitive solution. Changes to the CSR binary standard could disrupt the parser in the future. Moreover, large-scale, submodel-level access to data stored in CSR files could be a more broadly helpful feature to the thermal community and something worth incorporating in a future release of OpenTD. Further, since our implementation can omit the body data stored in the NODTRE file without data loss, incorporating this algorithm into OpenTD may offer memory savings to the broader TD community. Such steps would open avenues for more processing tools for large datasets and streamline the workflow for many TD users within the aerospace community.

## Acknowledgement

The research was carried out at the Jet Propulsion Laboratory, California Institute of Technology, under a contract with the National Aeronautics and Space Administration (80NM0018D0004).